\documentclass[pra,showpacs,floatfix,amsmath,amsfonts,twocolumn]{revtex4} 
\usepackage{graphics}
\usepackage{epsfig}
\usepackage{amssymb}
\usepackage{color}

\usepackage{amsthm,amscd,amsfonts,amssymb,epsfig,graphicx}

\begin{document}
\date{\today}

\title{Resonant scattering of matter wave gap-solitons by optical lattice defects}

\author{Valeriy A. Brazhnyi$^1$ and Mario Salerno$^2$}

\address{
$^{1}$Centro de F\'{\i}sica do Porto, Faculdade de Ci\^encias, Universidade do Porto, R. Campo Alegre 687, Porto 4169-007, Portugal\\
$^{2}$ Dipartimento di Fisica "E. R. Caianiello", CNISM and INFM Gruppo Collegato di Salerno,
Universit\'{a} di Salerno, via Ponte don Melillo I-84084, Fisciano (SA), Italy
}

\begin{abstract}
The physical mechanism underlying scattering properties of matter wave gap-solitons by linear optical lattice defects is investigated.
The occurrence of repeated reflection, transmission and trapping regions for increasing  strengths of an  optical lattice defect are shown to be due to  impurity modes inside the defect potential with chemical potentials and numbers of atoms matching corresponding quantities  of an incoming gap-soliton. For gap-solitons with chemical potentials very close to band edges,  the number of  resonances observed in the scattering coincides with  the number of bound states which can exist in the defect potential for the given defect strength.  The dependence of the positions  and widths of the transmission resonant on the incoming gap-soliton velocities are investigated by means of a defect mode analysis and effective mass theory. The comparisons  with direct  integrations of the Gross-Pitaevskii equation provide a very good agreement confirming the correctness of our interpretation.
  The possibility of multiple resonant transmission through arrays of optical lattice defects is also demonstrated. In particular, we show that it is possible to  design the strength of the defects so to balance the velocity detunings and to allow the resonant transmission through a larger number of defects. The possibility of using these results for very precise gap-soliton dynamical filters is suggested.
\end{abstract}


 \maketitle

\section{Introduction}

Bose-Einstein condensates (BECs) in optical lattices (OLs)  are presently attracting a great deal of interest~\cite{reviews} due both to their flexibility in parameter design and to the possibility they offer to  observe interesting phenomena such as superfluid to Mott-insulator transition~\cite{nature2002}, Bloch oscillations \cite{BO}, Landau-Zener tunneling~\cite{arimondo2005, arimondo2009}, generation of coherent atomic pulses (atom laser)~\cite{atom-laser}, atom interferometry~\cite{Fattori2008}, etc. In this respect, OLs allow to control important properties of BEC  by means of their periodic structure, this  allowing, for example,  the existence and  stability of localized nonlinear excitations with chemical potentials inside  band-gaps (so called  gap-solitons (GSs)) even in the presence of repulsive interactions (positive scattering lengths). This fact, that would be obviously  impossible in absence of the OL \cite{KS}, has been experimentally demonstrated in~\cite{GapSol}.

Modulations of  the OL can be used to accelerate, decelerate or to scatter GSs as well as to control their velocities~\cite{BKK}. Uniform accelerations of the OL combined with periodic modulations of the scattering length, either in space or in time, were shown to be effective tools to induce  long--lived Bloch oscillations in the nonlinear regime~\cite{SKB08}, as well as band-gap tunneling  phenomena such as the Landau-Zener tunneling \cite{accel_OL}. Periodic time dependent OL accelerations were also used to achieve the dynamical localization of nonlinear matter waves~\cite{BKS_EPL09, BKS_JPB09} and the Rabi-oscillations of GS states across a band-gaps~\cite{BKS_PRA09} which survive on a long time scale in the presence of nonlinearity. Moreover,  the combination of  the above  phenomena permits the stirring of GSs both in the reciprocal  and in the direct lattice space as recently demonstrated in  \cite{BKS_PRA10}. Besides the one-dimensional contexts in which these effects  have been investigated,  OLs also play an important role for  stabilizing GSs against collapse or decay in higher dimensions~\cite{BKS02}.

All what said above  refers to the case of perfect OLs, e.g. OLs  without distortions or defects which compromise the periodicity structure. In this case the interplay between periodicity and nonlinearity is the only source for the localization of matter waves in the system. OL defects, however,  introduce additional bound states  in the band-gaps (so called impurity modes) \cite{kivshar2001} providing  an alternate source  of localization in the system.  Opposite to GSs,  impurity modes exist both in absence and in presence of nonlinearity and can interfere in  the scattering process of GSs by OL defects.

Scattering properties  of solitons with single (non periodic) potential wells have been extensively investigated during  the past years. For the case of  the nonlinear Schr\"odinger equation (NLSE)  scattering of solitons by extended defects were  numerically investigated  in \cite{Stoychev04} where the occurrence of a series of reflection, transmission, and trapping regions as a function of the defect strength was reported.
A similar problem has been recently investigated  for solitons of the  Gross-Pitaevskii equation (GPE) with rectangular potential wells and for attractive interatomic interactions \cite{brandt2010}.

Similar studies were also done for point defects of the discrete NLSE  equation \cite{Wadati1992, Hasegawa1996, Primatarowa2005, Molina2006} which corresponds, under suitable conditions, to a tight--binding model of BEC in a deep OL \cite{smerzi}. The resonant transmission, reflection and trapping of discrete breathers by point-defects was investigated in \cite{Flach2003}. In contrast to these studies, however,  the scattering of continuous GPE GSs by OL defects have been scarcely investigated. In this context, we mention the numerical study performed in \cite{Ahufinger2007} where the  scattering properties of GSs  were suggested to be useful to construct  quantum switches and quantum memories.
To our knowledge, however, the mechanism underlying resonant transmissions of GPE GSs by OL defects and the possibility of multiple defects resonant scattering  have not yet been discussed.

In the present paper we provide an extensive  numerical investigation of  the scattering properties of  GSs by OL defects and identify the physical  mechanism underlying the phenomenon  of the  resonant transmission. In particular, we show that the presence  of  repeated reflections transmission and trapping regions observed for increasing  strengths of an  OL defect is associated to the impurity modes inside the defect potential with energies (chemical potentials) and numbers of atoms matching corresponding quantities  of the incoming GS. As the  OL defect strength is increased, new impurity modes enter
from the gap edges, moving toward the center of the gap (bottom of the potential). This implies that for incoming  GSs with chemical potentials very close to band edges,  the number of  resonances observed in the scattering coincides with  the number of bound states which can exist in the defect potential for the given defect strength. This fact is demonstrated both by studying   stationary states inside the defect potential and by direct numerical integrations of the GPE. An excellent agreement between the two approaches is found, this  confirming the correctness of our interpretation.

The dependence of the resonant transmission on the incoming velocity of the GS is also investigated both for fundamental GSs  in the semi-infinite gap and for GSs in the first gap zone.  As a result we show  that the transmission resonant peaks become wider as the incoming velocity is increased, with  very  sharp peaks at small  velocities. The multiple resonant transmission through a series of (two and three)  OL defects is also demonstrated. We show in this case that for equally spaced identical  OL defects the widths of the full transmission resonances in general decreases as the number of defects is increased. We demonstrate, however, that the resonant transmission through a series of defects can be achieved if defects are designed  so to compensate off-resonance detunings introduced by velocity changes. This fact gives rise to the possibility of using  arrays of OL defects as very precise filters for  matter wave dynamics.

The paper is organized as follows. In Sec. II we present  the model equation and discuss the basic properties of GSs of the GPE. In the Sec. III we consider  the interaction of small amplitude  GSs with a  localized Gaussian impurity in the OL. The problem is investigated by means of direct numerical integrations of the GPE for both attractive and repulsive interactions as well as for attractive and repulsive defects. In section IV we use a stationary defect mode  analysis  to show that the repeated resonant transmission, reflection and trapping regions  occurs in correspondence of resonances with impurity modes inside the defect potential and investigate their dependence on GS incoming velocities.   In section V the multiple  resonant transmission across a series of equidistant (equal and unequal) OL defects, is investigated. In Sect. VI the  main results of the paper are shortly resumed.

\section{Model equation and stationary localized solutions}

Let us consider a cigar-shaped BEC  described by the following normalized  one-dimensional GPE~\cite{Victor,reviews}
\begin{equation}
\label{GP}
i\frac{\partial\psi}{\partial t} = -\psi_{xx}+V_{ext}(x) \psi + \sigma |\psi|^{2}\psi,
\end{equation}
where $V_{ext}(x)$ denotes an external potential of the form: $V_{ext}(x)=V_{ol}(x)+ V_d(x)$, with  $V_{ol}(x)$ a perfect OL  of period  $L$:  $V_{ol}(x)=V_{ol}(x+L)$,  and $V_d(x)$ a defect potential consisting of a sum of $n_d$ single wells potentials  localized on a distance of several lattice period  around the OL sites $x_i$, $i=1,...,n_d$.
In the following  we assume $V_{ol}$ and  $V_{d}$  to have the form
\begin{eqnarray}
\label{PerPot}
&& V_{ol}(x)=V_0 \cos(2 x), \\
&& V_d(x)= \sum_{i=1}^{n_d} \frac{\eta_i}{\sqrt{2\pi}\Delta_i}\exp\left[-(x-x_i)^2/(2\Delta_i^2)\right]	 \label{defect}
\end{eqnarray}
where without loss of generality the period of the OL is taken $\pi$. We remark that  results of this paper will  not  qualitatively dependent on type of defect (we used Gaussian defects just for  numerical convenience) and similar results can be obtained for other shapes of the defects, like square well defects, for example).
In the following we will mainly restrict  to the cases of few OL defects:  $n_d= 1, 2, 3,$ only.
We also remark that in Eq.(\ref{GP}) the normalization has been made by measuring the energy  in units of recoil energy $E_r =\hbar^2 k^2 / (2m)$,  where $k=\pi/d$ and $d$ is the lattice constant, the  space coordinate and time in units of $d/\pi$ and $E_r/\hbar$, respectively. The dimensionless macroscopic wave function is also normalized as $\int |\psi|^2dx = 8\pi Nk|a_s|$, where $a_s$ is the s-wave scattering length.
\begin{figure}[ht]
\epsfig{file=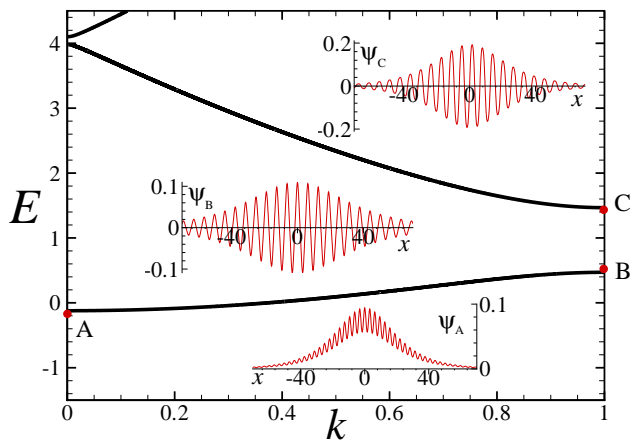,width=9cm}
\caption{Band-gap structure for $V_0=-1$ and GS bound states  at points $A$, $B$ and $C$ near band-gap edges, corresponding to chemical potentials $E_{A,B,C}=-0.125; 0.475; 1.452$, respectively. Insets show the  wave functions $\psi_{A,B,C}$ of the bound states.}
\label{fig_e(k)}
\end{figure}

It is well known that in the absence of the defect potential $V_d\equiv 0$, Eq.(\ref{GP})  posses families of exact GS solutions with energy located in the band-gaps of the linear eigenvalue  problem
\begin{equation}
\label{eigenval}
\frac{d^2 \varphi_{\alpha k}}{d x^2} + \left[ E_{\alpha}(k) - V_{ol}(x)\right]\varphi_{\alpha k}=0,
\end{equation}
where $\varphi_{\alpha k}(x)$ are orthonormal set of Bloch functions with $\alpha$ denoting the band index and $k$ the crystal-momentum  inside the first Brillouin zone (BZ): $k\in [-1,1]$. It is also known  that small-amplitude GSs with chemical potentials $E_s$ very close to band edges are of the form $\psi(x,t)=A(\zeta,\tau)\varphi_{\alpha k}(x)e^{-i E_{\alpha}(k)t}$ with the envelope function $A(\zeta,\tau)$ obeying the following NLSE
\begin{eqnarray}
i\frac{\partial A}{\partial \tau} = -\frac{1}{2M_{eff}} \frac{\partial^2 A }{\partial \zeta^2} + \chi |A|^2A \label{eq:env}
\end{eqnarray}
where  $\tau$ and $\zeta$ are slow temporal and spatial variables,  $M_{eff}=(d^2E_\alpha/dk^2)^{-1}$ denotes   the soliton effective mass and $\chi=\sigma\int |\varphi_{\alpha k}|^4 dx $ the effective nonlinearity \cite{KS}.
The condition for the existence of such solitons is  $\chi M_{eff} <0$ and coincides with the condition for the modulational instability of the Bloch wavefunctions at the edges of the BZ \cite{KS}. Examples  of small amplitude  GSs  with chemical potential inside the band-gap structure are depicted in Fig.  \ref{fig_e(k)}.

In the presence of very diluted  OL  defects, GSs  will continue to exist and away from defects they practically coincide with GSs of the undistorted OL. An attractive (local potential well) or repulsive (local potential barrier) OL defect will be seen by the GSs differently, depending on the sign of their effective mass. Thus, for example, in the case of repulsive interactions and a negative effective mass,  a GS approaching a repulsive defect will see it as a trapping potential (rather than as a potential barrier), thus besides being totally or partially transmitted/reflected, it can also be trapped at the defect site, a fact which would be impossible in absence of the OL.
In all the numerical simulations presented in  this paper  we have used stationary GSs of Eqs. (\ref{GP})--(\ref{defect}), exactly determined by the shooting method \cite{Alfimov2002} or by self-consistent calculations \cite{MSLPHYS},  and have  put them  in action  by means of phase imprinting (e.g. we multiply the state by the phase factor $e^{-i\sigma vx/2}$, with  $v$ being the GS velocity).
\begin{figure}[ht]
\centerline{
\epsfig{file=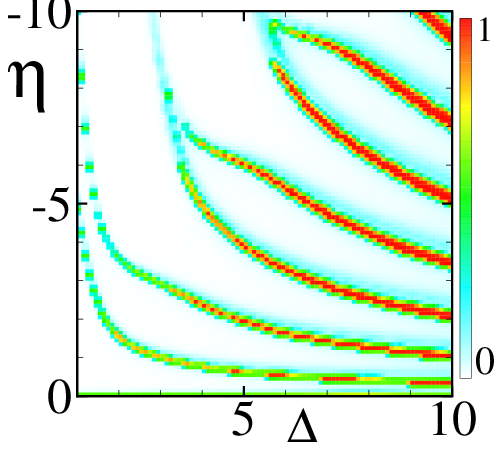,width=3.cm}
\epsfig{file=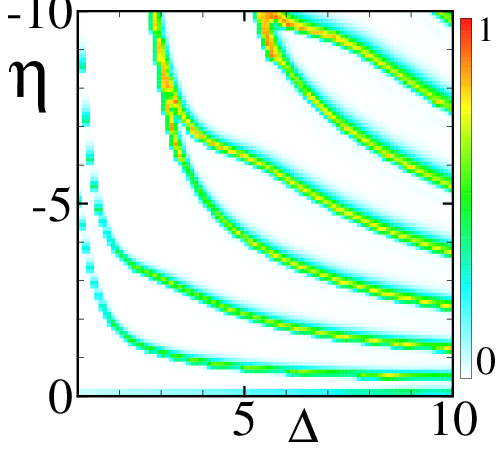,width=3.cm}
\epsfig{file=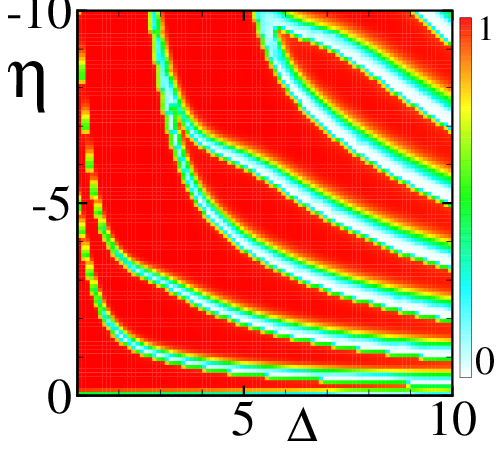,width=3.cm}}
\caption{Dependence of the transmission (left), trapping (center) and reflection (right) coefficients  on the amplitude $\eta$ and width $\Delta$ of the defect potential  $V_d$. Parameters are $v=0.05$, $E_s=-0.125$, $V_0=-1$.}
\label{eta(D)_w-0125_v005}
\end{figure}
This provides initial condition for  the GPE  numerical time integration of the form: $\psi(x, 0)= \psi_s(x) e^{-i\sigma vx/2}$. For possible experimental  implementations of our results, an alternate method to use to put the GS in action could be the acceleration of the OL for a short time interval to move the stationary state away from the BZ edges (center) so that it can acquire a small Bloch velocity $v_B=dE_\alpha(k)/dk\ll 1$ (for details on how this can be done see \cite{BKS_PRA10}).

\section{Resonant scattering of GS by an OL defect}

We consider first the scattering of a  GS by a  single  localized defect ($n_d=1$ in Eq. (\ref{defect})). In order  not to perturb the soliton initially, the distance between the soliton center and the OL defect is taken much larger than the  width of the GS ($\approx 80 L$).
In the following we compute the trapping, transmission and reflection coefficients defined as $C=N_C(t_f)/N_{ini}$, $T=N_T(t_f)/N_{ini}$ and $R=1-(T+C)$,  with  $N_C(t_f)=\int_{-x_c}^{x_c}|\psi(t_f)|^2 dx$, $N_T(t_f)=\int_{x_c}^{\infty}|\psi(t_f)|^2 dx$ and $N_{ini}=\int_{-\infty}^{\infty}|\psi(t=0)|^2 dx $ denoting the numbers of atoms trapped, transmitted and in the initial state, respectively. The final time $t_f$ depends on the initial velocity of the GS and in the numerical experiment is determined  as the time necessary for the coefficients $T,R,C$ to become stationary after the scattering process has occurred.
The trapping region $[-x_{c};x_{c}]$ has the  size of the initial soliton and in all our calculations we  fix  $x_{c}=30L$.

\subsection{Scattering of semi-infinite GS by an OL defect }

For the scattering of a GS in the semi-infinite gap to occur, the existence criterion for a  GS  near the bottom edge of the first band (where  the  effective mass is positive) implies that  the nonlinear coefficient  must be negative  sign$(\sigma)=-1$.
\begin{figure}[ht]
\epsfig{file=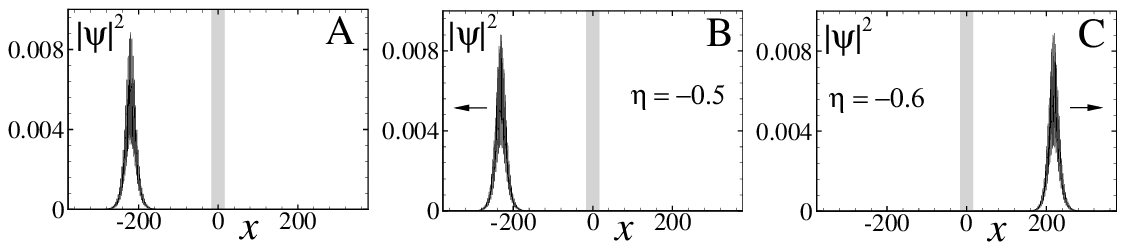,width=9cm}
\epsfig{file=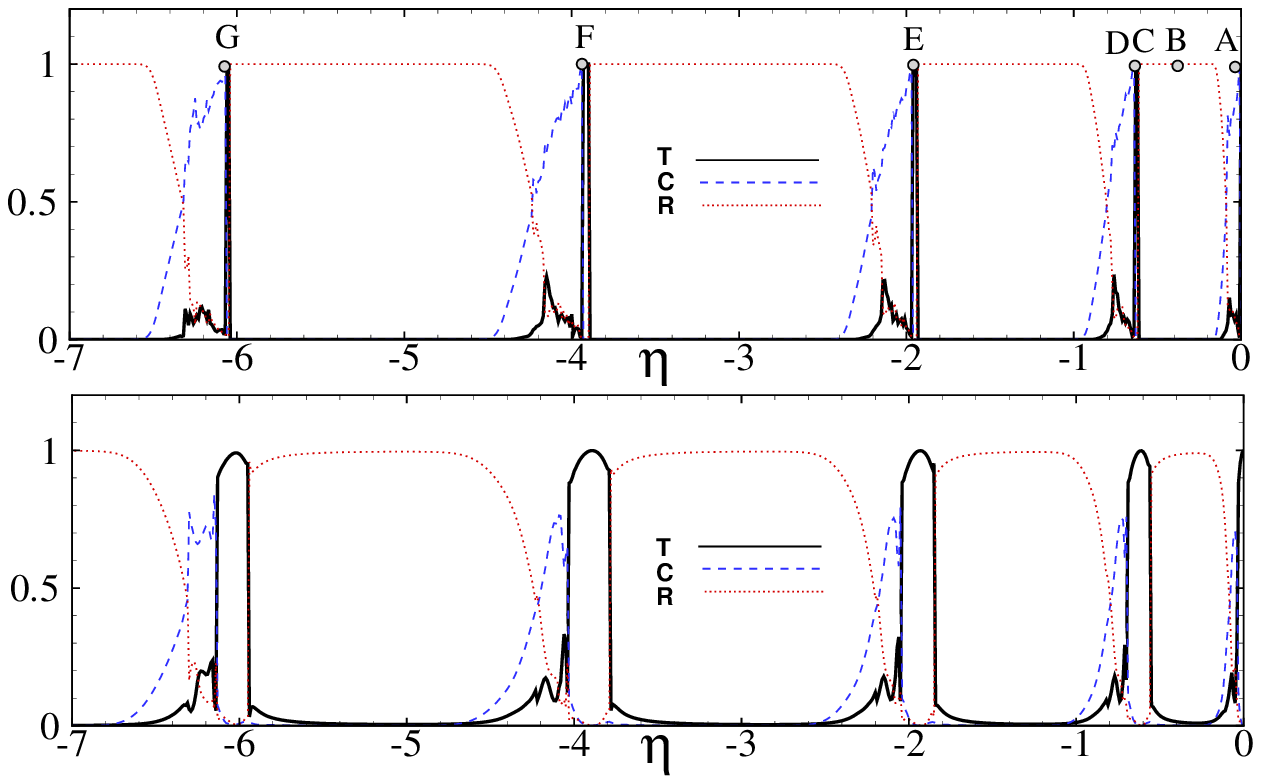,width=9cm}
\caption{
Dependence of coefficients $T(\eta)$, $R(\eta)$ and $C(\eta)$ on the OL defect strength
for two GS incoming velocities: $v=0.01$ (middle panel)
and $v=0.05$ (bottom panel).
The top panels A, B and C correspond to the profiles of
initial, reflected and transmitted GS. The wide line in gray shows
the position of the defect. Other parameters are $\Delta=5$,
$E_s=-0.125$, $V_0=-1$.}
\label{fig3}
\end{figure}
By applying a small  initial velocity to the GS in the defect direction, depending on the amplitude $\eta$ and width $\Delta$ of the defect,  three possible scenarios can occur: i) complete reflection, $R=1$; ii) complete transmission, $T=1$; iii)  partial trapping, $C>0$.
The regions of the parameter space $\{\eta, \Delta\}$ where these different regimes occur, as obtained from direct numerical integration of Eq.(\ref{GP}),  are reported  in Fig. \ref{eta(D)_w-0125_v005}.
The dependence of $T, C, R$ on the defect strength $\eta$ for two different incoming GS velocities and for  a  fixed value of the defect width $\Delta=5$ is depicted in Fig. \ref{fig3}. We see that by changing the strength of the defect it is possible to achieve complete reflections ($R\approx 1$)  or transmission ($T\approx 1$) as well as partial  trappings ($C>0$). The profiles of the initial, reflected and transmitted GS are depicted in the top panels of  Fig.\ref{fig3} for defect strengths corresponding to points labeled in the middle panel by letters A, B, C. By comparing the middle  and the bottom panels of  Fig.\ref{fig3} it is clear that the sharp peaks at small incoming velocities ($v\approx 0$) for which $T\approx 1$ (see points A, C, E, F, G in the middle panel) become wider as the velocity is increased while the regions for which $R\approx 1$ are a bit reduced (also see Fig.\ref{fig3_transm_0.475} for the case of first band-gap GSs).

\begin{figure}[ht]
\epsfig{file=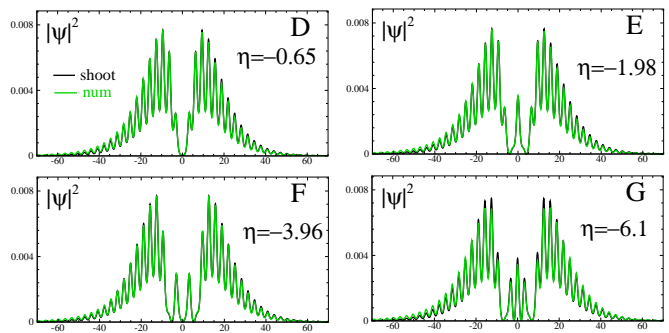,width=9cm}
\caption{Defect modes corresponding to the maxima of the trapping curves ($C\approx 1$) indicated in Fig.\ref{fig3} by letters D, E, F and G, respectively. Green lines correspond to the defect modes obtained from the dynamical GPE calculations while dark lines refer to defect mode profiles obtained from  Eq.(\ref{Matheiu_def}) (see below). Parameters are fixed as in Fig.\ref{fig3}.
}
\label{modes_fig3}
\end{figure}

The first four impurity modes corresponding to the transmission peaks $D, E, F, G$ in the middle panel of Fig.\ref{fig3} have been  depicted in Fig. \ref{modes_fig3}. Notice the alternating odd-even symmetry of these modes (modes D and F being odd  and modes E and G being even with respect to the center $x=0$ of the defect potential) as usual for eigenstates of one-dimensional trapping potentials.
We remark that the existence of four resonant transmission peaks (and reflection regions) seen in Fig.\ref{fig3} for $-7\le \eta < 0$ correlates with the existence of four impurity modes for the given defect strength region
(see stationary defect mode analysis below).

\subsection{Scattering of first band-gap GSs by an OL defect}
Similar results as those of the previous section can be found for GSs inside  the first band-gap, with the only  difference that now there are two possibilities for the existence of  small-amplitude solitons: i) in the vicinity of the top edge  of the first band where the effective mass is negative and therefore GSs can exist  only for repulsive interactions $\sigma=1$; ii)  and in the vicinity of the bottom edge of the second band where effective mass is positive and GS exist only for attractive interactions $\sigma=-1$. The small-amplitude GS near the first band-gap edges B,C, in  Fig.\ref{fig_e(k)}  are shown by the corresponding profiles  $\psi_B$ and $\psi_C$ depicted in the figure.

\begin{figure}[ht]
\centerline{
\epsfig{file=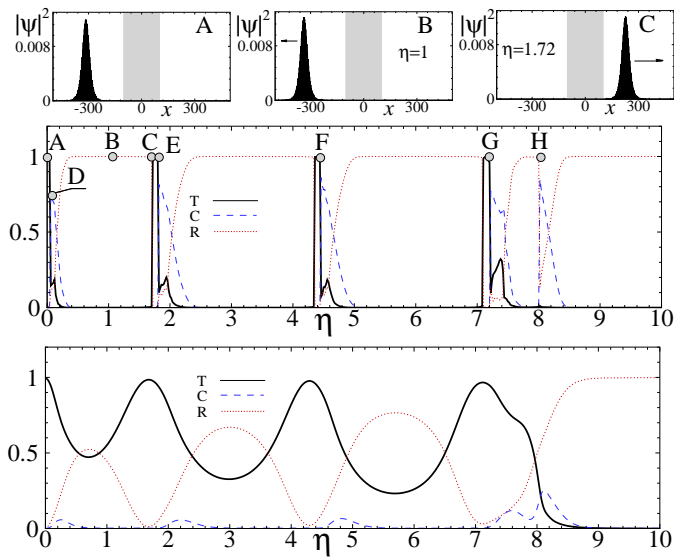,width=9cm}}
\caption{Dependence of the transmission, reflection and trapping coefficients on the defect amplitude $\eta$ for $\Delta=5$ and for two different incoming velocities of the GS ($E_s=0.475$): $v=0.02$ (middle panel)  and $v=0.1$ (bottom panel). Top panels A-C show the initial, reflected and transmitted GS profiles, respectively.
}
\label{fig3_transm_0.475}
\end{figure}

\begin{figure}[ht]
\centerline{
\epsfig{file=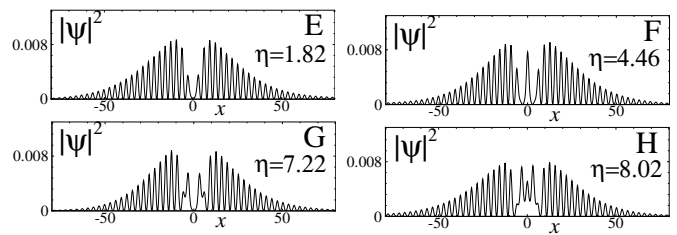,width=9cm}}
\caption{First four defect modes (panels E-H) in correspondence of maxima of the C$(\eta)$ curve in Fig. \ref{fig3_transm_0.475}. Parameters are the same as in Fig. \ref{fig3_transm_0.475}.
}
\label{fig3_transm_0.475_modes}
\end{figure}

\begin{figure}[ht]
\centerline{
\epsfig{file=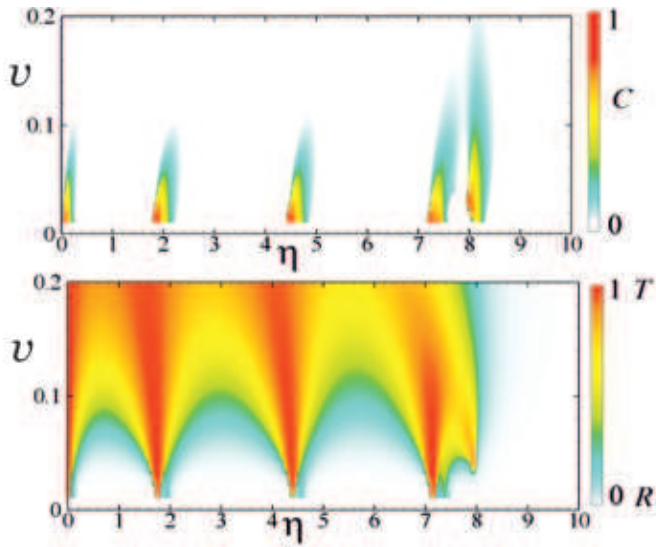,width=9cm}
}
\caption{Dependence of $C(\eta, v)$ (top panel) and  T$(\eta,v)$, R$(\eta,v)$ (bottom panel) on $\eta$. Parameters are fixed as:  $E_s=0.475$, $\Delta=5$, $V_0=-1$, $\sigma=1$.
}
\label{fig3_transm_0.475_v}
\end{figure}

As remarked before, the sign of the effective mass determines the type of the interaction of the GS has with the defect potential and for negative GS effective mass (repulsive interatomic interactions) the defect will be seen as a defect trapping potential (supporting therefore bound states) if the defect strength $\eta$ is positive rather than negative (as seen for the case of a positive effective mass). Except for this, results go in parallel with those of the previous section and have been collected  in Figs.\ref{fig3_transm_0.475}, \ref{fig3_transm_0.475_v} for the case of an initial GS close the top edge  of the first band 	(negative effective mass). In particular, from the bottom panel of Fig. \ref{fig3_transm_0.475} and Fig. \ref{fig3_transm_0.475_v} we clearly see that for a fixed  defect width the region of the resonant transmission (corresponding to red color) becomes wider  as the incoming GS velocity is increased, while the full reflected regions are achieved mainly for small velocities, as one could have expected. We remark that also in this case the number of resonant transmission reflection and trapping regions are found to correlate  with the number of impurity modes present in the defect potential for the given range of the defect strength (first four modes corresponding to the maxima of the trapping coefficient C in Fig.\ref{fig3_transm_0.475} are depicted in Fig.\ref{fig3_transm_0.475_modes}, \ref{modes_fig4_0475}).

\section{Impurity mode analysis and GS scattering by OL defects }
As remarked before, the presence of a  repulsive (attractive) OL defect in the GPE affects
the existing GS states for $\eta=0$ and introduces  additional localized states inside the gap
in presence of repulsive (attractive) nonlinearities.

Numerical calculations show that the band structure is only slightly affected by an OL impurity, the main effect being  the introduction  of bound states spatially localized at the impurity sites and with chemical potentials inside  the band-gap.
Note that
repulsive (attractive) OL defects in the presence of an attractive (repulsive) nonlinearity cannot introduce additional bound states  because the corresponding impurity potentials correspond  to   barriers rather than potential wells, due to the positive (negative) effective mass.
Recalling  that the opposite signs of nonlinearity and effective mass is a necessary condition for the GS existence, one has that the effective impurity potential acts as a trapping potential when $\eta$ and $\sigma$ have equal signs. Away from the OL impurity localized states  are practically the same as for $\eta=0$ case. At the defect site, however, GSs levels get  slightly shifted by the impurity potential and additional impurity modes enter the gap.
A GS moving through the impurity will  have, in general, a mismatch in energy and in number of atoms with the impurity modes, this giving a partial  reflection/transmission  of the incoming matter wave.

\begin{figure}[ht]
\epsfig{file=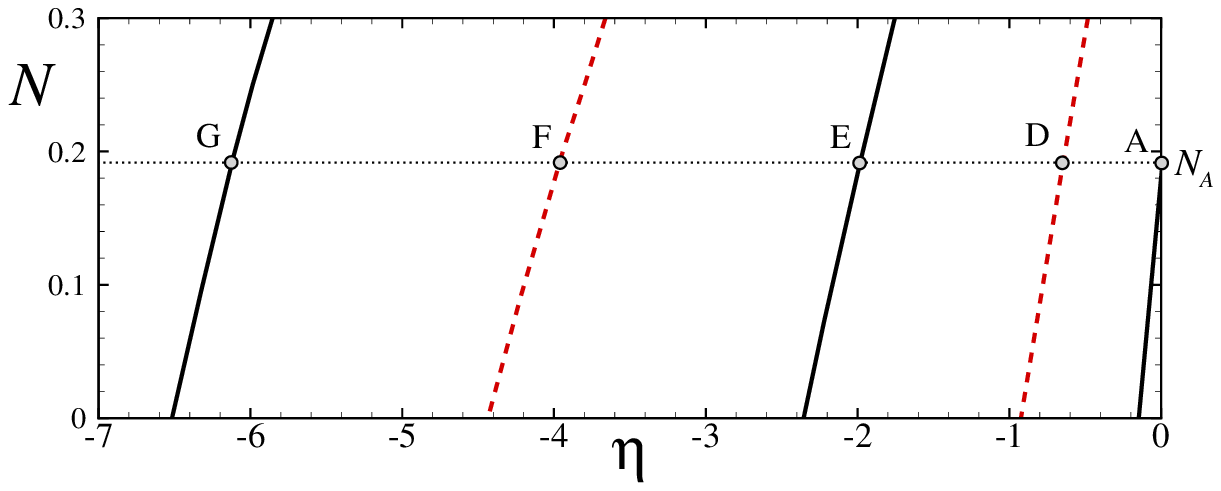,width=9cm}
\epsfig{file=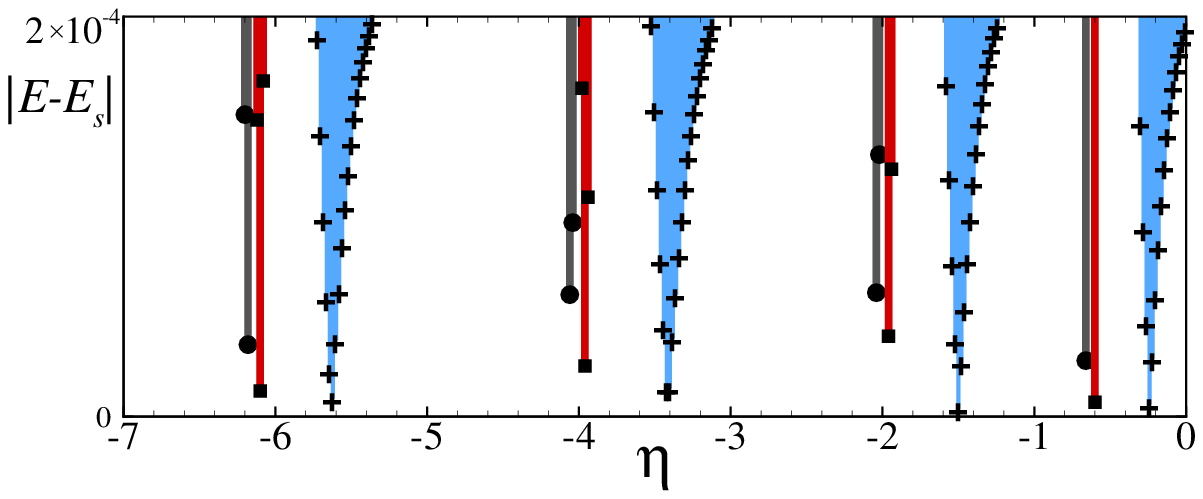,width=9cm}
\caption{Top panel. Number of atoms  in defect modes {\it vs} the amplitude of the defect $\eta$. The (black,solid)/(red,dashed) lines correspond to the even/odd symmetry of the defect modes.
Other parameters are $E_s=-0.125$, $\Delta=5$, $V_0=-1$, $\sigma=-1$.
Bottom panel. Mismatch  between defect and  GS energies, $|E-E_s|$, vs $\eta$ for different incoming velocities: $v\approx 0$ (gray, $\bullet$), $v=0.05$ (red, $\blacksquare$) and $v=0.1$ (blue, {\bf +}).
}
\label{modes_fig4_-0125}
\end{figure}

\begin{figure}[ht]
\epsfig{file=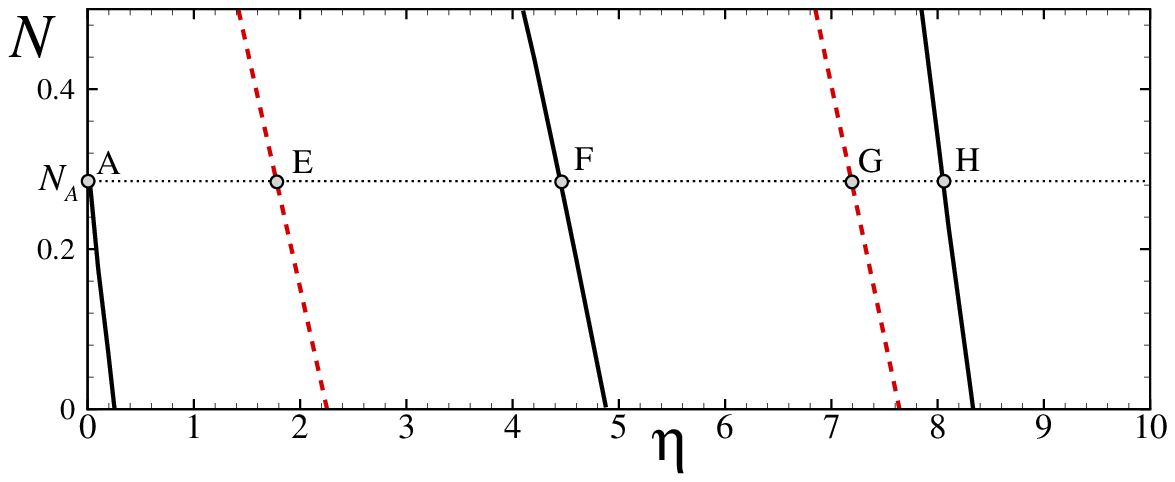,width=9cm}
\epsfig{file=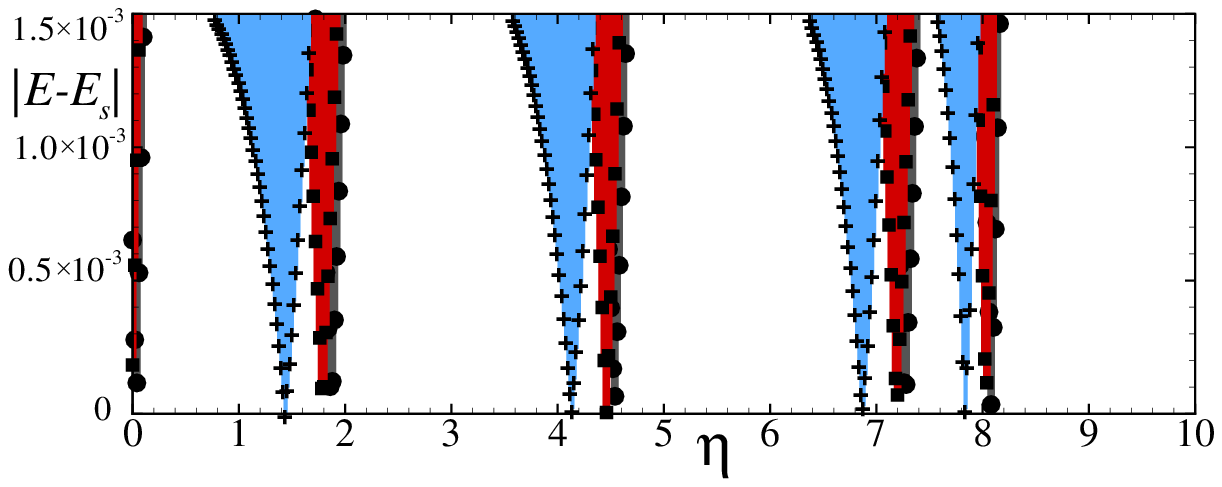,width=9cm}
\caption{The same as in Fig.\ref{modes_fig4_-0125} but for $E_s=0.475$ and $\sigma=1$. The incoming soliton velocities are $v\approx 0$ (gray, $\bullet$), $v=0.1$ (red, $\blacksquare$) and $v=0.2$ (blue, {\bf +}). }
\label{modes_fig4_0475}
\end{figure}

A total transparency of the OL defect is expected for
incoming GS energies and number of atoms exactly matching those of an impurity modes inside the defect  potential (notice that the energy of impurity modes depend on $\eta$ and on the number of atoms).
By increasing the strength of the impurity, the depth of the defect  potential increases and more impurity modes  enter the gap. As $|\eta|$ is increased, the energies of these
modes enter the gap from the top (bottom) of a band for impurity strength and the nonlinearity both positive (negative). This implies that  the  transparency (complete transmission) of the impurity occurs in correspondence of each impurity mode entering the gap and matching
the energy of the incoming GS given by
\begin{equation}
E\approx E_s^{(0)} + \frac 12 M_{eff}\, v_B^2,
\label{energyshift}
\end{equation}
where the first term $E_s^{(0)}$ represents the energy of the stationary GS state at $k=k_0$ (eg. with $v_B(k_0)=0$), while the second one is  the contribution due to the kinetic energy (here $v_B(k)=dE_\alpha (k)/dk$ is the Bloch velocity  and $M_{eff}=(d^2E_\alpha/dk^2)^{-1}|_{k=k_0}$ the effective mass).  For small velocities
the energy (\ref{energyshift}) practically coincides with $E_s^{(0)}$ but in general the kinetic term should be  accounted in the matching in energy with the impurity levels (see below).
Notice that Eq. (\ref{energyshift}) is only valid near stationary points $k_0$ (bottoms or tops of a band) where $v_B(k)\approx (k-k_0)/M_{eff}$ and $E$ can be written as
\begin{equation}
E \approx E_s^{(0)} + \frac{(k-k_0)^2}{2  M_{eff}}.
\label{energyshiftapprox}
\end{equation}
In the range of initial velocities $v_B=[0, 0.2]$ we have considered, the energy curves $E_1(k)$ in vicinity of $k_0=0$ and $k_0=1$ are very well approximated by Eq. (\ref{energyshiftapprox}) with $E_1(0)=-0.12177$,  $E_1(1)=0.47065$, and $M_{eff}\approx 0.565$ and $M_{eff}\approx-0.167$ for bottom and top edges of the band, respectively.	
%
By knowing $v_B$ (e.g. $k$) and $M_{eff}$ one can compute the energy shift due to the nonzero velocity to be accounted in the matching between the GS and the impurity levels (see lower panels of Figs. \ref{modes_fig4_-0125}, \ref{modes_fig4_0475}).
Notice that the kinetic energy has the sign of the effective mass so that $E$ is pushed forward the corresponding band edge for finite $v_B$, meaning that  inside the impurity potential the GS matching condition with an impurity mode can be achieved for a lower values of $|\eta|$. From this we expect  the resonance transmission peaks to be shifted away from the $v=0$ resonance
toward lower values of $|\eta|$ as $v_B$ is increased.

\begin{figure}[ht]
\centerline{
\epsfig{file=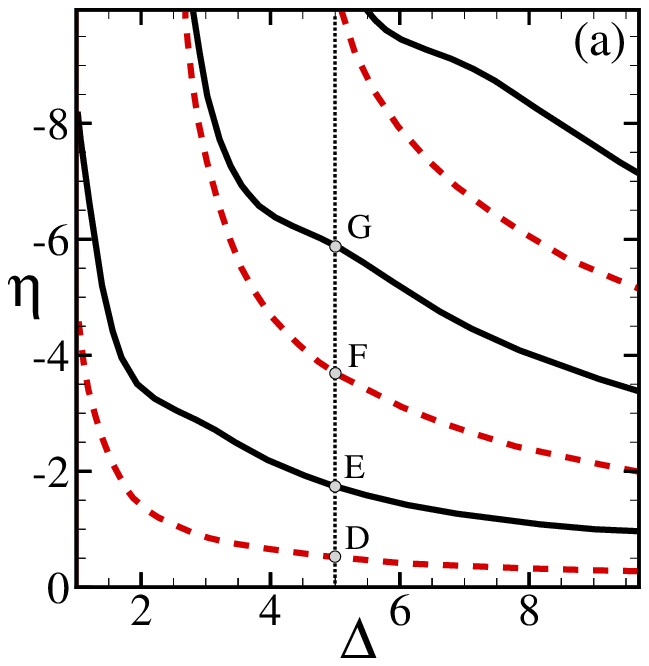,width=4.5cm}
\epsfig{file=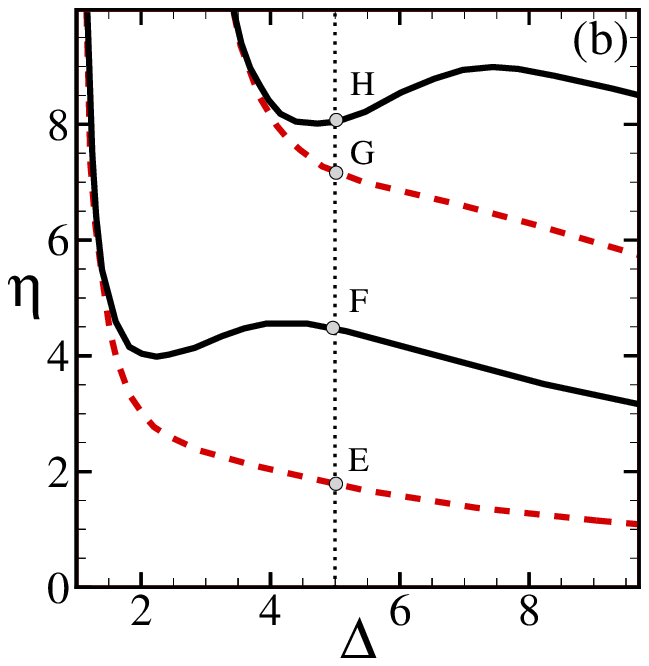,width=4.5cm}}
\caption{Level curves $N=N_A$ for even (red, dashed) and odd (black, solid) defect modes {\it vs} $(\eta, \Delta)$ for the case of semi-infinite gap (a) and bottom of the first gap (b).
Other parameters are fixed as $E_s=-0.125(0.475)$, $V_0=-1$, $\sigma=-1(1)$.
}
\label{-0125_odd_even_eta(D)}
\end{figure}

To confirm these prediction with GPE calculations we have solved the stationary problem
\begin{eqnarray}
	\label{Matheiu_def}
	u_{xx}+[E-V_{ol}(x)-V_{d}(x)]u-\sigma u^3=0
\end{eqnarray}
searching for bound states $u(x)$ localized at the OL defect using both  shooting method and
self-consistent calculations. Using these approaches we found that the values of $\eta$ for which the Gaussian defect becomes transparent to the GS dynamics, correspond to the values for which a localized mode inside the defect potential  has energy and number of atoms matching the corresponding quantities of the incoming GS. Since the discussion is very similar (a part implications due to signs  of the effective masses) for GS of the semi-infinite and for the ones inside finite-gaps,  we refer to the case of an initial GS inside the semi-infinite gap, with a positive effective mass with chemical potential close to the bottom of the first band.
In the top panel of  Fig.\ref{modes_fig4_-0125} we have shown the dependence of the number of atoms $N$ in a defect mode of a given symmetry on $\eta$,  for an energy $E=E_s=-0.125$ corresponding to a GS with zero incoming velocity (point A of Figs. \ref{fig_e(k)}, \ref{fig3}). The horizontal dotted line refers  to the numbers of atoms in the initial GS. In the bottom panel of Fig. \ref{modes_fig4_-0125} we show the  energy mismatch $|E-E_s|$ between the energy of a defect modes $E$  and the one of an incoming GS given by  Eq. {\ref{energyshiftapprox}} for different incoming velocities. Notice that the intersection points $A, D, E, F, G$ of the dotted line $N=N_A$ with the curves $N(\eta)$ are in coincidence with the zeros of the function $|E-E_s|$ displayed in the bottom panel for  $v\approx 0$ and correspond to the maxima of the transmission coefficient $T(\eta)$ in  Fig.\ref{fig3} (see curves $v\approx 0$ in the middle panel).
 Fig. \ref{modes_fig4_-0125} (see also \ref{modes_fig4_0475} for repulsive case) also explains the decay of the reflection coefficient and the rising of the trapping coefficient coinciding up to the maxima observed (just before the resonance)  for increasing values of $|\eta|$ away from the resonance points $A, D, E, F, G$. To the right of these points in  Fig.\ref{modes_fig4_-0125}, the number of particles in the defect mode is higher then the number of particles in the incoming soliton $N(\eta)>N_A$. In this case the GS cannot "use" a  defect mode to pass the OL impurity and will be fully reflected.
 On the left of the resonance points, the number of atoms in the GS is higher than the one in the defect mode so that the GS can be captured by the OL impurity by releasing the excess number of atoms into the reflection and transmission channels. It is clear that the peak of the capture coefficient occurs just before  the resonant transmission (e.g. for $0<N_A-N(\eta)\ll 1$) of the GS through the defect (achieved when the condition $N(\eta)=N_A$ is exactly fulfilled). Notice from Fig.\ref{modes_fig3} that at the resonant transmission, the profiles of the stationary impurity  modes obtained by solving Eq.(\ref{Matheiu_def}) exactly coincide with the solution of the GPE during the passage trough the defect. In Fig.\ref{-0125_odd_even_eta(D)} we have also depicted, similarly to Fig.\ref{eta(D)_w-0125_v005}, the level curves $N=N_A$ as a function of the amplitude $\eta$ and the width $\Delta$ of the defect.
From the bottom panel of Fig. \ref{modes_fig4_-0125} it is also clear that for $v \neq 0$ the resonances shift in the direction of lower values of $|\eta|$ (as expected from our analysis).
 This well  correlates  with the GPE calculations reported in the bottom panel of  Fig.\ref{fig3}. Similar results are  obtained for a GS near the  bottom of the first band gap (point B in Fig. \ref{fig_e(k)}) for the case of repulsive interactions. This is shown in   Fig.\ref{modes_fig4_0475}. from which we see that  the shift of the resonance due to the finite Bloch incoming velocity is always in the direction of lower values of $\eta$ and are in  good agreement with the GPE  numerical results reported in the bottom panel of Fig. \ref{fig3_transm_0.475}.
In particular, for the velocity $v_B=0.1$ and  the resonance near $\eta=4$ we obtain from Fig. \ref{modes_fig4_0475} that the resonance peak is at $\eta\approx 4.45$ while from the GPE result in Fig. \ref{fig3_transm_0.475} we obtain the value $\eta\approx 4.3$.
It is also worth to note from the bottom panels of Figs. \ref{modes_fig4_-0125}, \ref{modes_fig4_0475}, that for a fixed energy mismatch the widths of the curves increase as the incoming velocity is increased, a fact which correlates with the broadening of the transmission peaks observed in the GPE calculations (compare bottom panels in Fig. \ref{fig3} and in Fig \ref{fig3_transm_0.475}, respectively).

From this we conclude that the above impurity mode analysis fully confirms the interpretation of the transmission peaks as resonances between incoming GSs  and impurity modes.
\begin{figure}[ht]
\centerline{
\epsfig{file=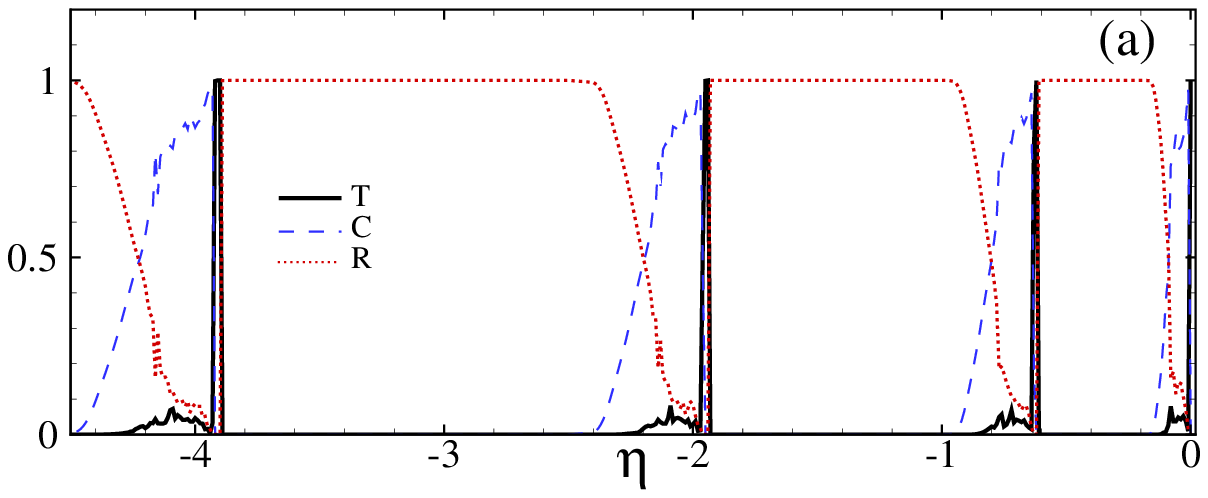,width=9.cm}
}
\centerline{
\epsfig{file=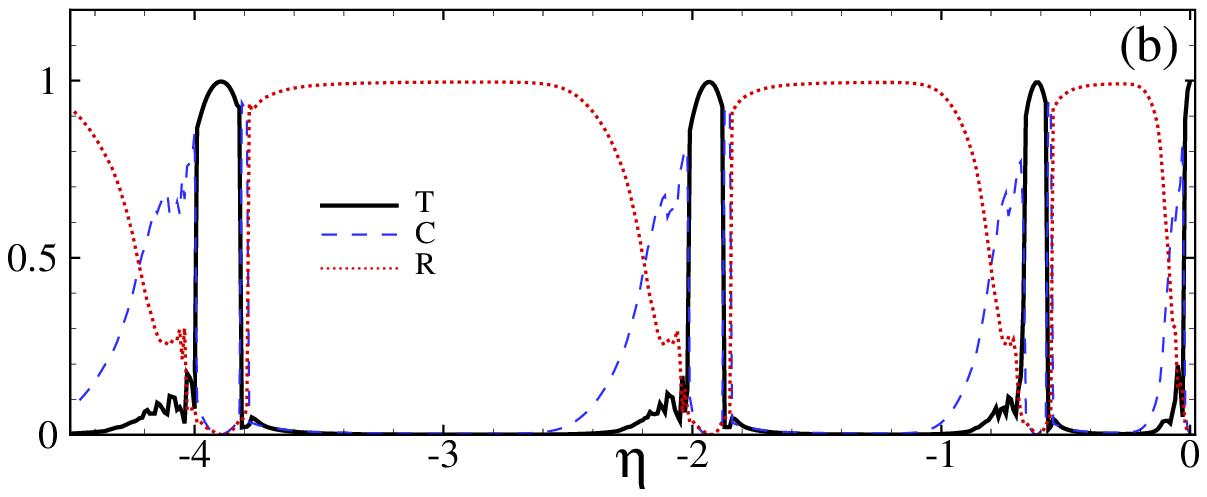,width=9.cm}
}
\caption{
Transmission, reflection and trapping diagram of GSs with velocities $v=0.01$ (a) and $v=0.05$ (b) in presence of two Gaussian defects of the OL placed at positions $x=0$ and $x=100\pi$. Other parameters are the same as in Fig.\ref{fig3}.}
\label{fig10}
\end{figure}
\section{Resonant transmission through multiple OL defects}
An interesting question to ask is whether  the soliton resonant transmission could survive multiple  impurity scatterings.   In Fig. \ref{fig10} we show  the transmission, reflection and trapping regions of an attractive GS of the GPE with  two identical Gaussian OL impurities, placed at $x=0$ and $x=100\pi$, respectively.
The TCR curves depend also on the distance between  the two impurities and this could  be varied so to find
optimal values  for  double resonant transmission to occur.
The phenomenon, however, is more sensitive to variations of the incoming GS velocity as one can see by comparing Fig.\ref{fig10}  with the case of single impurity in Fig. \ref{fig3}.
\begin{figure}[ht]
\centerline{
\epsfig{file=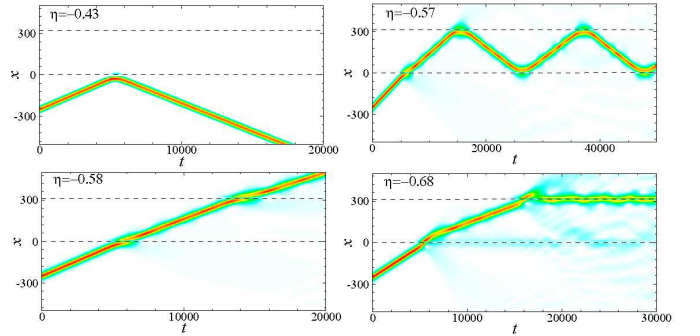,width=9.cm}
}
\caption{Contour plot of the GS  space-time dynamics for an OL with two equal Gaussian defects of strengths:
$\eta=-0.43$ (top left), $\eta=-0.57$ (top right), $\eta=-0.58$ (bottom left), $ \eta=-0.68$ (bottom right). The initial velocity of the GSs is $v=0.05$ and  parameters are the same as for Fig. \ref{fig10} (bottom panel).}
\label{fig11}
\end{figure}
Notice that  while the transmission resonance peaks at small velocities are practically unaffected by the presence of the second impurity, they become more narrow at larger velocities for the resonant scattering on two impurities.
The shrinking of the resonances at larger velocities can be understood by the fact that for higher incoming velocities  the velocity of the GS after the first impurity is slightly reduced and the variation introduces a detuning from resonance in the scattering with the second impurity which in turn reduces the double resonance width.
From this one can expect that in the presence of more lattice  defects the multiple resonance transmission peaks become very narrow and only solitons with very precise initial matching velocities will be able to pass.
In general the GS may be able to pass only a finite number of impurities before remaining trapped at one impurity or becoming scattered back and forth between them. This is shown in Fig.\ref{fig11} where contour plots of the GS  space-time dynamics in presence of two identical Gaussian defects are reported for different strengths $\eta$ and for the same parameters as for the bottom panel of Fig.\ref{fig10}.
In the left top and bottom panels we see the occurrence of total reflection and transmission for  a resonant value of $\eta$  while in the corresponding right panels we show  the back and forth dynamics of the soliton between  two impurities (top panel) and  the trapping   by the second impurity (top panel) for a non resonant values of $\eta$. Notice the small change of the soliton velocity after the passage through the first defect.
To overcome the detuning from resonance induced by the change of velocity one could  design the strength of the second impurity so to  match the value of the
intermediate velocity and still achieving $T\approx 1$. In this way one can realize a double (or multiple) filtering of the soliton motion so that only GSs having
very precise initial velocity can overcome  the defect series.
\begin{figure}[ht]
\centerline{
\epsfig{file=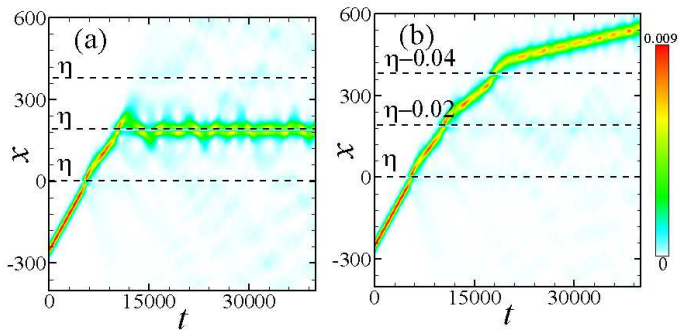,width=9.cm}
}
\caption{
Panel (a). Contour plot of the GS  space-time dynamics in an OL with three equal Gaussian defects  at positions $x_1=0$, $x_2=60\pi$, $x_3=120\pi$ of strengths $\eta_1=\eta_2=\eta_3=-0.67$. Panel (b). Case of three different Gaussian defects of strengths $\eta_1 =-0.67$, $\eta_2 =\eta_1-0.02$, $\eta_3=\eta_1-0.04$ (b). Other parameters are fixed as in Fig.\ref{fig3}.}
\label{fig12}
\end{figure}
This possibility is illustrated in Fig. \ref{fig12} where the resonant transmission of a GS though a series of three Gaussian defects of the OL, with defect strengths designed so to compensate the detunings introduced by the velocity changes, is shown. Notice from the left panel of Fig. \ref{fig12} that in the case of equal defects with the initial velocity matching the resonance transmission peak of the first defect, the GS cannot  be transmitted through all the three defects but it is stopped at the second defect. The dynamics of the soliton in presence of multiple defects may be quite complicated and  is beyond  the aim of this work (an investigation in the parameter space of the possible  scenarios for the GS time evolution will be discussed elsewhere).

\section{Conclusion}

In this paper  we have investigated the scattering properties of  matter wave gap-solitons with  optical lattice defects in the framework of the mean-field Gross-Pitaevskii equation.
 We have shown that the occurrence of repeated reflection, transmission and trapping regions  are in correspondence of the defects strengths for which  the number of atoms and energies of additional bound states created by the optical lattice defect,  match the ones of the incoming gap-soliton. This has been demonstrated by a study of the stationary defect modes energies (chemical potentials) and number of atoms as a function of the defect strength.  A very good agreement between  the predicted values of the resonant transmission peaks by means  of impurity modes  and the ones found by direct time integrations of the Gross-Pitaevskii equation, is found.

 The behavior of the  reflection and trapping curves have also been explained by impurity mode analysis.  These investigations have been  performed  both for attractive and repulsive interactions and for localized states both in the semi-infinite gap (attractive case) and in the first gap zone (attractive and repulsive cases). The dependence of the resonant transmission on the initial gap-soliton velocity has been also investigated. We have shown that the positions  of resonant transmission peaks shift toward lower values $|\eta|$ as the initial  gap-soliton velocity is increased, while the  widths of resonances  shrinks to zero  at very small gap-soliton velocities. The possibility of multiple resonant transmission through an arrays of defects was also demonstrated. In particular, we have shown that for an optical lattice with two equally spaced identical  Gaussian  defects the widths of the full transmission resonances for larger incoming velocities, decreases as the number of defects is increased. The sharpening of the transmission peaks has been explained in terms of the  detuning from resonance introduced by the small velocity change after the passage of an optical lattice defect. Finally, the resonant transmission of a gap-soliton though a series of  Gaussian defects  with unequal  strengths designed so to compensate detunings  introduced by the velocity changes, was demonstrated. These results give the possibility to construct very precise
filters for the matter wave gap-soliton dynamics by means of properly designed arrays of OL defects.

\section*{Acknowledgments}
V.A.B. acknowledges support from the FCT grant, PTDC/FIS/64647/2006.
MS acknowledges the MIUR (PRIN-2008 initiative) for partial financial support.


\begin{thebibliography}{9}

\bibitem{reviews} See review papers O. Morsch and M. Oberthaler, Rev. Mod. Phys. {\bf 78}, 179 (2006) and references therein; I. Bloch, J. Phys. B {\bf 38}, S629. (2005); D. Jaksch, and P. Zoller, Ann. Phys. N.Y. {\bf 315}, 52 (2005); V.A. Brazhnyi and V. V. Konotop, Mod. Phys. Lett. {\bf 18} 627 (2004).

\bibitem{nature2002} M. Greiner, O. Mandel, T. Esslinger, T.W. H\"ansch, and I. Bloch,
Nature (London) {\bf 415}, 39 (2002).

\bibitem{BO}O. Morsch, J. H. M\"uller, M. Cristiani, D. Ciampini, and
E. Arimondo, Phys. Rev. Lett. {\bf 87}, 140402 (2001); I. Carusotto, L. Pitaevskii, S. Stringari, G. Modugno, and M. Inguscio, Phys. Rev. Lett. {\bf 95}, 093202 (2005).

\bibitem{arimondo2005} M. Jona-Lasinio, O. Morsch, M. Cristiani, N. Malossi, J. H. M\"uller, E. Courtade, M. Anderlini, and E. Arimondo, Phys. Rev. Lett. {\bf 91}, 230406 (2003); S. Wimberger, R. Mannella, O. Morsch, E. Arimondo, A. R. Kolovsky, and A. Buchleitner,
Phys. Rev. A {\bf 72}, 063610 (2005).

\bibitem{arimondo2009} A. Zenesini, H. Lignier, G. Tayebirad, J. Radogostowicz,
D. Ciampini, R. Mannella, S. Wimberger, O. Morsch, and
E. Arimondo, Phys. Rev. Lett. {\bf 103}, 090403 (2009).

\bibitem{atom-laser} B. P. Anderson and M. A. Kasevich, Science {\bf 282}, 1686 (1998).

\bibitem{Fattori2008}
M. Fattori, C. D'Errico, G. Roati, M. Zaccanti, M. Jona-Lasinio, M. Modugno, M. Inguscio, and G. Modugno,  Phys. Rev. Lett. {\bf 100}, 080405 (2008).

\bibitem{KS} V. V. Konotop and M. Salerno, Phys. Rev. A {\bf 65}, 021602 (2002).


\bibitem{GapSol} B. Eiermann, Th. Anker, M.Albiez, M. Taglieber, P. Treutlein, K. P. Marzlin, and M. K. Oberthaler, Phys. Rev. Lett. {\bf 92}, 230401 (2004).

\bibitem{BKK}  V.A. Brazhnyi, V.V. Konotop, V. Kuzmiak, Phys. Rev. A {\bf 70}, 043604 (2004).

\bibitem{SKB08} M. Salerno, V. V. Konotop, and Yu. V. Bludov, Phys. Rev. Lett. {\bf 101}, 030405 (2008).

\bibitem{accel_OL} V. V. Konotop, P. G. Kevrekidis, and M. Salerno, Phys. Rev. A {\bf 72}, 023611 (2005);
V.A. Brazhnyi, V.V. Konotop, V. Kuzmiak, V.S. Shchesnovich, Phys. Rev. A {\bf 76}, 023608 (2007).

\bibitem{BKS_EPL09} Yu. V. Bludov, V. V. Konotop, and M. Salerno, Europhys. Lett.  {\bf 87}, 20004 (2009).

\bibitem{BKS_JPB09} Yu. V. Bludov, V. V. Konotop, and M. Salerno, J. Phys. B: At. Mol. Opt. Phys. {\bf 42} 105302 (2009).

\bibitem{BKS_PRA09} Yu. V. Bludov, V. V. Konotop, and M. Salerno, Phys. Rev. A {\bf 80}, 023623 (2009).

\bibitem{BKS_PRA10} Yu.V. Bludov, V.V. Konotop, and M. Salerno, Phys. Rev. A {\bf 81} 053614 (2010).

\bibitem{BKS02} B.B. Baizakov, V.V. Konotop, and M. Salerno, J. Phys. B {\bf 35} 5105 (2002);
E.A. Ostrovskaya, Yu.S. Kivshar, Phys. Rev. Lett. {\bf 90}, 160407 (2003); J. Yang and Z. Musslimani, Opt. Lett. {\bf 23}, 2094 (2003); B.B. Baizakov, B.A. Malomed, and M. Salerno, Europhys. Lett. {\bf 63}, 642 (2003).

\bibitem{kivshar2001}A.A. Sukhorukov, Y.S. Kivshar, O. Bang, J.J. Rasmussen, and P.L. Christiansen, Phys. Rev. E {\bf 63}, 036601 (2001).

\bibitem{Stoychev04} K.T. Stoychev, M.T. Primatarowa, and R.S. Kamburova, Phys. Rev. E {\bf 70}, 066622 (2004).

\bibitem{brandt2010} T. Ernst and J. Brand, Phys. Rev. A {\bf 81}, 033614 (2010).


\bibitem{Wadati1992}T. Iizuka and M. Wadati, J. Phys. Sot. Jpn. {\bf 61}, 4344 (1992).

\bibitem{Hasegawa1996}T. Iizuka, H. Amie, T. Hasegawa, C. Matsuoka, Phys. Lett. A {\bf 220}, 97 (1996).

\bibitem{Primatarowa2005}M.T. Primatarowa, K.T. Stoychev, and R.S. Kamburova, Phys. Rev. E {\bf 72}, 036608 (2005).

\bibitem{Molina2006} Luis Morales-Molina and Rodrigo A. Vicencio, Opt. Lett.  {\bf 31}, 966 (2006).

\bibitem{smerzi}A. Trombettoni and A. Smerzi, Phys. Rev. Lett. {\bf 86}, 2353 (2001); F.K. Abdullaev, B.B. Baizakov, S.A. Darmanyan, V.V. Konotop, and M. Salerno, Phys. Rev. A {\bf 64}, 043606 (2001).

\bibitem{Flach2003}A. E. Miroshnichenko, S. Flach, and B. Malomed, Chaos {\bf 13},  874 (2003).

\bibitem{Ahufinger2007}  V. Ahufinger, A. Mebrahtu, R. Corbal\'an and A. Sanpera, New Journal of Physics {\bf 9}, 4 (2007).

\bibitem{Victor} V. M. P\'erez-Garc\'{\i}a, H. Michinel and H. Herrero, Phys. Rev. A {\bf 57}, 3837 (1998).

\bibitem{Alfimov2002}  G. L. Alfimov, V. V. Konotop and M. Salerno, Europhys. Lett. {\bf 58}, 7 (2002).

\bibitem{MSLPHYS} M. Salerno, Laser Physics {\bf 15}, 620 (2005).

\end{thebibliography}
\end{document}